# DANCING BUNCHES AS VAN KAMPEN MODES

A. Burov, FNAL*, Batavia, IL 60510, U.S.A.

## Abstract

Van Kampen modes are eigen-modes of Jeans-Vlasov equation [1-3]. Their spectrum consists of continuous and, possibly, discrete parts. Onset of a discrete van Kampen mode means emergence of a coherent mode without any Landau damping; thus, even a tiny couple-bunch wake is sufficient to drive instability. Longitudinal instabilities observed at Tevatron [4], RHIC [5] and SPS [6] can be explained as loss of Landau damping (LLD), which is shown here to happen at fairly low impedances. For repulsive wakes and single-harmonic RF, LLD is found to be extremely sensitive to steepness of the bunch distribution function at small amplitudes. Based on that, a method of beam stabilization is suggested.

## OIDE-YOKOYA EXPANSION

For stability analysis, the steady state distribution function, or the phase space density $F(I)$, can be treated as an input function, determined either by cooling-diffusion kinetics, or by injection. After the steady state problem being solved as in Ref. [7], the phase space density perturbation $f(I,\varphi,t)$ is to be found from the Boltzmann-Jeans-Vlasov (BJV) equation [8]:

$$\frac{\partial f}{\partial t} + \Omega(I)\frac{\partial f}{\partial \varphi} - \frac{\partial V}{\partial \varphi}F'(I) = 0 \,. \quad (1)$$

Here $I$, $\varphi$ and $\Omega(I)$ are the canonical action, phase, and incoherent frequency in the distorted potential well;

$$V(z,t) = -\int f(I',\varphi',t)W(z-z')dI'd\varphi' \quad (2)$$

is a perturbation of a single-particle Hamiltonian by wake fields induced by the beam perturbation $f(I,\varphi,t)$ with the wake function $W(z)$ [9]. Following Oide and Yokoya [10], the eigenfunctions of Eq. (1) may be expanded in Fourier series over the phase:

$$f(I,\varphi,t) = e^{-i\omega t}\sum_{m=1}^{\infty}\left[f_m(I)\cos m\varphi + g_m(I)\sin m\varphi\right] \quad (3)$$

With $z(I,\varphi=0) = \min_\varphi z(I,\varphi)$, this yields an equation for the amplitudes $f_m(I)$:

$$\left[\omega^2 - m^2\Omega^2(I)\right]f_m(I) = $$
$$-2m^2\Omega(I)F'(I)\sum_{n=1}^{\infty}\int dI' V_{mn}(I,I')f_n(I'). \quad (4)$$

The matrix elements

$$V_{mn}(I,I') = -\frac{2}{\pi}\int_0^\pi d\varphi\int_0^\pi d\varphi'\cos(m\varphi)\cos(n\varphi')\times \quad (5)$$
$$\times W(z(I,\varphi)-z(I',\varphi'))$$

can be expressed in terms of the impedance $Z(q)$ [9]:

$$W(z) = -i\int_{-\infty}^{\infty}\frac{dq}{2\pi}\frac{Z(q)}{q}\exp(iqz), \quad (6)$$

yielding

$$V_{mn}(I,I') = -2\,\mathrm{Im}\int_0^\infty dq\,\frac{Z(q)}{q}G_m(q,I)G_n^*(q,I'); \quad (7)$$
$$G_m(q,I) \equiv \int_0^\pi \frac{d\varphi}{\pi}\cos(m\varphi)\exp[iqz(I,\varphi)].$$

Note that bunch-to-bunch interaction is neglected here. Equation (4) reduces the integro-differential BJV equation (1) to a standard eigen-system problem of linear algebra.

Hereafter, dimensionless units are used. Dimensionless action $I$ is converged to conventional eV·s units with a factor of $E_0\Omega_0/(\eta\omega_{\mathrm{rf}}^2)$, where $E_0 = \gamma mc^2$ is the beam energy, $\eta = \gamma_t^{-2} - \gamma^{-2}$ – the slippage factor, $\gamma$ – relativistic factor, $\omega_{\mathrm{rf}}$ – RF angular frequency, and $\Omega_0$ is zero-amplitude incoherent synchrotron frequency in a bare RF potential. For single-harmonic RF, considered in this paper, bucket acceptance (maximal action) in the dimensionless units is $8/\pi\approx 2.54$. Dimensionless synchrotron frequencies are measured in units of $\Omega_0$. The dimensionless impedance $Z(q)$, Eq. (7), relates to the conventional $Z_\parallel(q)$ as $Z(q)=DZ_\parallel(q)$ with the intensity factor $D = Nr_0\eta c\omega_{\mathrm{rf}}^2/(\Omega_0^2\gamma C)$, where $N$ is the bunch population, $r_0$ - the classical radius and $C$ – the machine circumference. The offset $z$ is RF phase in radians.

## VAN KAMPEN MODES

More than half a century ago, N. G. van Kampen found an eigensystem of Jeans-Vlasov equation for infinite plasma [1-3]. In general, a spectrum of these modes consists of continuous and discrete parts. The continuous spectrum essentially describes single-particle motion, accompanied with a proper plasma response. Frequency band of the continuous spectrum is one of the incoherent frequencies. Continuous modes are described by singular functions, underlying their primary relation to single-particle motion. Landau damping of coherent motion can be treated as phase mixing of the continuous van Kampen modes. As opposed to the continuous spectrum, the discrete one not necessarily exists. Discrete modes are described by regular functions; some, if not all, of them do not decay with time. Indeed, since the original equations (analogue of Eq. (4)) have real coefficients, the mode frequencies are either real or forming complex-conjugate pairs. So, it is either a pure loss of Landau damping or instability. Infinite plasma with monotonic distribution density was shown to be always stable; discrete modes of the pure LLD type may only appear if the distribution function is of a finite width. Plasma instability was shown to be possible for non-monotonic distributions only [11].

For bunch longitudinal motion, eigenproblem of Jeans-Vlasov equation was first considered by A. N. Lebedev [12]. Although the suggested formalism was not numerically tractable, an important result was analytically obtained: it was proved that for the space charge impedance above transition, all the eigen-frequencies are always real.

For pure parabolic RF potential, van Kampen modes were analyzed for various wake functions [10, 13-15]. For that model RF, rigid bunch oscillations at the unperturbed synchrotron frequency is always a solution of equation of motion [13]. Indeed, single-particle equations of motion can be written as

$$\ddot{z}_i + z_i = \sum_j W'(z_i - z_j); \quad i, j = 1, ..., N.$$

The solution can be presented as a sum of a steady-state-related part $\hat{z}_i$ and a small perturbation $\tilde{z}_i$. It is clear that rigid-bunch motion $\tilde{z}_i = \text{const} \cdot \cos t$ satisfies this equation. While that rigid-bunch frequency is intensity-independent, all the incoherent frequencies are typically either suppressed or elevated by the potential well distortion; thus, normally this mode stays outside the incoherent band, so it is a discrete LLD type mode. This normally expected behavior is not necessarily though. As it was shown in Ref [13], for broad band impedance model, core and tail incoherent frequencies may go with intensity in opposite directions, so the rigid-bunch mode may be covered by incoherent frequencies; thus, it should be Landau-damped in this case. In Ref. [16], loss of Landau damping was analyzed for the space charge impedance and various RF shapes, assuming it is the rigid-bunch mode which is losing its Landau damping. That assumption is not correct when the RF frequency spread is taken into account: action dependence of the emerging discrete mode is normally very different from the rigid-bunch one [7]. Because of that, rigid bunch approximation overestimates the threshold intensity. For the space charge and the Hofmann-Pedersen distribution, this overestimation is ~3 times below transition and ~5 times above it. However, above transition the LLD is not radical [7], so it is very sensitive to tiny tails of the distribution.

The analysis is limited hereafter by the weak head-tail approximation for the dipole azimuthal mode: only $m=n=1$ matrix elements are left in Eq.(4).

## INDUCTIVE IMPEDANCE

Hadron machines are normally dominated by the resistive wall or inductive impedances. In the dimensionless units, inductive wake function and impedance are written as

$$W(s) = -k\delta(s); \quad Z(q) = -ikq;$$
$$k = -\frac{2Nr_0\eta\omega_{\rm rf}^3}{\gamma c \Omega_0^2} \frac{{\rm Im}\,Z_\|}{nZ_0}, \quad (8)$$

where $Z_0 = 4\pi/c = 377$ Ohm, and $n$ is the revolution harmonic number.

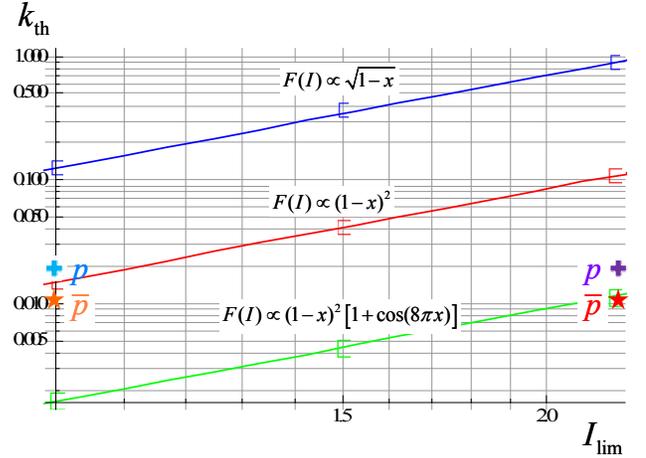

Figure 1: Threshold LLD intensity parameter $k_{\rm th}$ versus the bunch emittance for 3 denoted distributions $F(I)$, where $x = I/I_{\rm lim}$. Lines are fits with $k_{\rm th} \propto I_{\rm lim}^{5/2}$. Tevatron data for protons (crosses) and pbars (stars) are shown at injection (right) and top energy (left symbols).

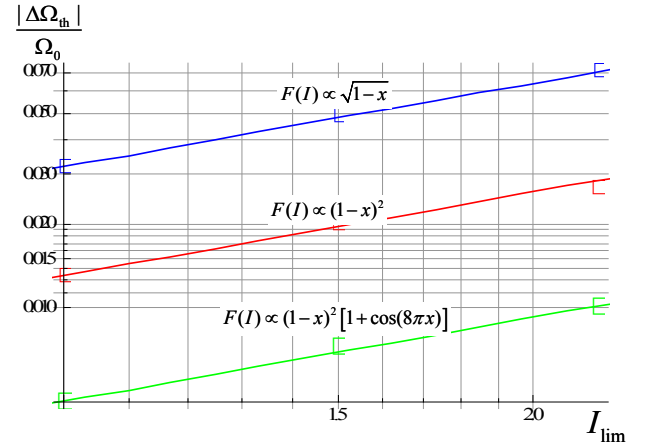

Figure 2: Same thresholds as above, in terms of the zero-amplitude relative incoherent frequency shifts found as in Ref. [7]. Lines are linear fits.

LLD threshold lines, $k_{\rm th}$ versus $I_{\rm lim}$, are presented in Fig. (1) for $k>0$ (repulsive wake: inductance above, or space charge below transition) and three distribution functions, with $x \equiv I/I_{\rm lim} \leq 1$: $F(I) \propto (1-x)^{1/2}$ (most stable), $F(I) \propto (1-x)^2$ (medium stable), and $F(I) \sim (1-x)^2(1+\cos(8\pi x))$ (least stable). The last distribution simulates a coalesced proton bunch in the Tevatron. It takes about an hour for memory of the constituent 7 bunches to get smeared in the coalesced proton bunch in the Tevatron.

Although the emergent discrete modes are far from being similar to the rigid-bunch motion $\propto \sqrt{I} F'(I)$, the power law $k_{th} \propto I_{lim}^{5/2}$ agrees with the prediction of Ref. [4, 17]. According to that, LLD happens when the zero-amplitude synchrotron tune shift $\Delta\Omega = \Omega(0) - \Omega_0 \propto \mathrm{Im}\, Z(l^{-1})/l^2 \propto k/I_{lim}^{3/2}$ exceeds the synchrotron tune spread in the nonlinear RF, $\delta\Omega \propto I_{lim}$. Equating these two values, one gets the threshold scaling $k_{th} \propto I_{lim}^{5/2}$ for the inductive impedance, confirmed by Figs. (1,2). The plots show strong dependences of the threshold intensity on details of the distribution function. Qualitatively, this can be interpreted as a high sensitivity to the distribution steepness at small arguments. That high sensitivity should not be too surprising. While wakes are stronger for shorter bunches, the frequency spread is weaker for them. That is why a small central portion of a bunch is less stable than the entire bunch. Effective length of the oscillating bunch centre depends on the distribution function: the steeper is the distribution at small amplitudes, the shorter this part is. This prediction appears to be generally correct when the bare RF synchrotron frequency monotonically decreases with the amplitude, and for any effectively repulsive wake field, diminishing the incoherent synchrotron frequencies.

At the Tevatron, long-living oscillations of proton bunches are seen both at the injection and top energy, if the damper is off; the oscillations are conventionally called as 'dancing bunches' [4]. For the antiprotons, similar phenomenon is observed at collisions only. For the Tevatron impedance model [18], the proton bunches at injection turn out to be ~2 times above the green line LLD threshold of Fig. (1). At the top energy, they are ~20 times above that threshold, and slightly above the red line threshold. According to these calculations, the antiprotons stay at the green line threshold at injection, and they are 10 times above it at the top energy. In reality, they are stable at injection, and unstable at collisions. To conclude, both proton and antiproton stability observations are in a reasonable agreement with the model described.

Since LLD strongly depends on the small-argument steepness of the distribution function, its local flattening there should increase the threshold. This sort of flattening can be achieved by means of RF phase modulation at a narrow frequency band around zero-amplitude synchrotron frequency. This RF shaking should smear distribution for the low-action resonant particles and thus make the bunch more stable. Dedicated experiments with that RF shaking were performed at the Tevatron. Observations are confirming; their detailed description is presented in a special report [19].

## CONCLUSIONS

Emergence of a discrete van Kampen mode means either loss of Landau damping or instability. Longitudinal bunch stability is analysed in weak head-tail approximation for inductive impedance and single-harmonic RF.

The LLD threshold intensities are found to be rather low: for cases under study all of them do not exceed a few percent of the zero-amplitude incoherent synchrotron frequency shift, strongly decreasing for shorter bunches. Because of that, LLD can explain longitudinal instabilities happened at fairly low impedances at Tevatron [4], and possibly for RHIC [5] and SPS [6], being in that sense an alternative to the soliton explanation [5, 20]. Although LLD itself results in many cases in emergence of a mode with zero growth rate, any couple-bunch (and sometimes multi-turn) wake would drive instability for that mode, however small this wake is. LLD is similar to a loss of immune system of a living cell, when any microbe becomes fatal for it.

The emerging discrete mode is normally very different from the rigid-bunch motion; thus the rigid-mode model significantly overestimates the LLD threshold. The power low of LLD predicted in Ref. [17] agrees with results of this paper. However, the numerical factor in that scaling low strongly depends on the bunch distribution function. Particularly, for inductive impedance above transition and three examined distributions, the highest LLD threshold intensity exceeds the lowest one by a factor of ~100. Based on that observation, proper RF phase shaking as a method of beam stabilization is suggested.

## REFERENCES

[1] N. G. van Kampen, Physica (Utrecht) **21**, 949 (1955).
[2] N. G. van Kampen, Physica (Utrecht) **23**, 647 (1957).
[3] G. Ecker, "Theory of Fully Ionized Plasmas", Academic Press, 1972.
[4] R. Moore et al., "Longitudinal bunch dynamics in the Tevatron", Proc. PAC 2003, p. 1751 (2003).
[5] M. Blaskiewicz et al.,"Longitudinal solitons in RHIC", Proc. PAC 2003, p. 3029.
[6] E. Shaposhnikova, "Cures for beam instabilities in CERN SPS", Proc. HB 2006, Tsukuba, 2006.
[7] A. Burov, "Van Kampen modes for bunch longitudinal motion", Proc. HB 2010, Switzerland.
[8] M. Henon, Astron. Astrophys., **144**, 211 (1982).
[9] A. Chao, "Physics of Collective Beam Instabilities", Whiley Interscience, 1993.
[10] K. Oide and K. Yokoya, KEK Preprint 90-10 (1990).
[11] O. Penrose, Phys. Fluids, **3**, 258 (1960).
[12] A. N. Lebedev, Atom. Energiya **25** (2), 100 (1968).
[13] M. D'yachkov and R. Baartman, Part. Acc. **50**, 105 (1995).
[14] K. Oide, Part. Acc, **51**, 43 (1995).
[15] Y. Shobuda and K. Hirata, Phys. Rev. E, **60**(2), 2414 (1999)
[16] O. Boine-Frankenheim and T. Shukla, Phys. Rev. ST-AB, **8**, 034201 (2005).
[17] V. Balbekov, S. Ivanov, IHEP preprint 91-14 (1991).
[18] Fermilab report, "Run II Handbook", p. 6.58 (1998).

[19] A. Burov and C-Y Tan, "Bucket shaking stops bunch dancing in Tevatron", Proc. PAC'11.
[20] M. Blaskiewicz et al., Phys. Rev. ST Accel. Beams **7**, 044402 (2004).